\documentclass[twocolumn]{aastex63}

\newcommand{\jwst}{JWST}
\newcommand{\hst}{HST}
\newcommand{\hz}{high-$z$}
\newcommand{\lya}{{Ly$\alpha$}}
\newcommand{\ha}{H$\alpha$}
\newcommand{\hi}{H\,{\sc i}}
\newcommand{\cii}{[C\,{\sc ii]}}
\newcommand{\buv}{$\beta$}
\newcommand{\muv}{$M_{\rm UV}$}
\newcommand{\ewlya}{EW$_0$(Ly$\alpha$)}
\newcommand{\fescLya}{$f_{\rm esc}^{\rm Ly\alpha}$}
\newcommand{\ksio}{$\xi_{\rm ion, 0}$}
\newcommand{\Llya}{$L$(\lya)}
\newcommand{\lgLya}{${\rm log}_{10}\,L$(\lya)}
\newcommand{\Dd}{$\Delta d_{\rm Ly\alpha}$}
\newcommand{\Dv}{$\Delta v_{\rm Ly\alpha}$}
\newcommand{\Reff}{$R_{\rm e}$}

\shorttitle{COSMOS-Web Imaging LAEs at $z\approx6$}
\shortauthors{Ning et al.}

\begin{document}

\title{Unveiling Luminous \lya\ Emitters at $z\approx6$ through JWST/NIRCam Imaging in the COSMOS Field}

\author[0000-0001-9442-1217]{Yuanhang Ning}
\altaffiliation{ningyhphy@mail.tsinghua.edu.cn}
\affiliation{Department of Astronomy, Tsinghua University, Beijing 100084, China}

\author[0000-0001-8467-6478]{Zheng Cai}
\affiliation{Department of Astronomy, Tsinghua University, Beijing 100084, China}

\author[0000-0001-6052-4234]{Xiaojing Lin}
\affiliation{Department of Astronomy, Tsinghua University, Beijing 100084, China}

\author[0000-0002-9634-2923]{Zhen-Ya Zheng}
\affiliation{CAS Key Laboratory for Research in Galaxies and Cosmology, Shanghai Astronomical Observatory, Shanghai 200030, China}

\author[0000-0003-0174-5920]{Xiaotong Feng}
\affiliation{Department of Astronomy, School of Physics, Peking University, Beijing 100871, China}
\affiliation{Kavli Institute for Astronomy and Astrophysics, Peking University, Beijing 100871, China}

\author[0000-0001-6251-649X]{Mingyu Li}
\affiliation{Department of Astronomy, Tsinghua University, Beijing 100084, China}

\author[0000-0002-3119-9003]{Qiong Li}
\affiliation{Jodrell Bank Centre for Astrophysics, University of Manchester, Oxford Road, Manchester M13 9PL, UK}

\author[0000-0002-9074-4833]{Daniele Spinoso}
\affiliation{Department of Astronomy, Tsinghua University, Beijing 100084, China}

\author[0000-0003-0111-8249]{Yunjing Wu}
\affiliation{Department of Astronomy, Tsinghua University, Beijing 100084, China}

\author[0000-0003-2273-9415]{Haibin Zhang}
\affiliation{Department of Astronomy, Tsinghua University, Beijing 100084, China}

\begin{abstract}
We study a sample of 14 spectroscopically confirmed \lya\ Emitters (LAEs) in the late era of reionization (at redshift $z\approx6$) based on the JWST/NIRCam imaging dataset. These LAEs with high \lya\ luminosity of \Llya~$\sim10^{42.4-43.4}$~erg~s$^{-1}$ have been covered by the (ongoing) COSMOS-Web survey \citep{cosmosWebb, casey23} over $0.28$ deg$^2$ in four NIRCam bands (F115W, F150W, F277W, and F444W). With JWST/NIRCam imaging, we determine the UV continua with \muv\ ranging from ${-}20.5$ to ${-}18.5$~mag. The UV slopes have a median value of \buv~$\approx-2.35$, and the steepest slopes can reach \buv~$<-3$. Under an excellent spatial resolution of JWST, we identify three out of the sample as potential merging/interacting systems. The 14 LAEs (and their components) are compact in morphology residing substantially below the mass-size relation of \hz\ galaxies. We further investigate their physical properties including the stellar mass ($M_*$) and star-formation rates (SFRs). Most of the LAEs lie on the SFR-$M_*$ main-sequence relation while two of them featured as ``little red dots" likely host active galactic nuclei (AGN), implying a ${\sim}10\%$ AGN fraction. Moreover, we reveal that a new correlation may exist between \lya\ equivalent width and the offset between \lya\ and UV emission (\Dd) with a median \Dd\ $\sim1$ kpc. This could be explained by \lya\ radiative transfer process in both ISM and CGM. The results usher a new era of detailed analysis on \hz\ LAEs with the JWST capability.
\end{abstract}


\section{Introduction}

Thanks to the {\it James Webb Space Telescope} (\jwst; \citealt{gardner06}), the exploration of high-redshift (\hz) galaxies has recently extended to $z\gtrsim9$, a highly neutral universe during the epoch of reionzation \citep[e.g.,][]{castellano22, treu22, adams23, bradley23, donnan23, furtak23, 2023arXiv230800751F}. Previous works can only use {\it Hubble Space Telescope} (\hst) to study $z\gtrsim6$ galaxies in detail with both relatively high resolution and sensitivity \citep[e.g.,][]{finkelstein12, jiang13a, jiang13b, bouwens14, schmidt16, tilvi16}. Such space observations usually provide faint samples of \hz\ objects in a small area. 

More luminous galaxies may own complex structures such as \lya\ blobs \citep[e.g.,][]{ouchi13, sobral19b, zhang20}. They are thus an ideal laboratory to study formation, interaction and evolution of galaxies. Currently, only large ground-based surveys can help us find most of luminous galaxies. With faint Lyman-break galaxies (LBGs) mostly found by \hst\ \citep[e.g.,][]{yan12, laporte15, oesch14, infante15, mcLeod16}, a large number of \hz\ bright LBGs are selected by the dropout technique from large-area ground-based observations \citep[e.g.,][]{laporte12, cl12, ono18}. The narrowband (or \lya) technique done by large-area ground-based observations provides a complementary method to find \hz\ galaxies which are relatively luminous \lya\ emitters \citep[LAEs; e.g.,][]{kashikawa06, kashikawa11, hu10, jiang17, zheng17, taylor20}.

The \hz\ galaxies (e.g., the LBGs and LAEs) have been analyzed based on the space telescopes such as \hst\ \citep[e.g.,][]{taniguchi09, jiang13a, jiang13b, Paulino-Afonso18, jiang20a} and {\it Spitzer} observations \citep[e.g.,][]{egami05, huangj11, labbe13, ashby15, bradac20, strait20, whitler23}. However, their physical properties have not been well constrained. For example, {\it Spitzer} covers the rest-frame optical bands for the \hz\ galaxies but its low resolution makes photometric measurements tentative for the distant small objects. The advent of \jwst\ are solving such problems with the transformative capabilities. Especially, the recently ongoing \jwst\ Cycle 1 program COSMOS-Web \citep[GO 1727;][]{cosmosWebb, casey23} will contribute the largest \jwst\ imaging survey area and cover a number of luminous \hz\ galaxies.

We have carried out a spectroscopic survey to build a large and homogeneous confirmed sample of \hz\ galaxies \citep{jiang17}. Besides constraining cosmic reionization using LAEs at $z\approx5.7$ and 6.6 \citep{ning20, ning21}, we also aim to study their galaxy properties in detail. As the survey are designed to select \hz\ galaxies from the famous fields, many of them will be covered by the \jwst\ observations including the above mentioned COSMOS-Web. The multiple JWST/NIRCam bands will reveal their individual properties on global and spatially resolved scales. Currently, the COSMOS-Web imaging survey has covered galaxies at $z~\sim~6$ which are spectroscopically confirmed by our M2FS survey. In this work, we focus on the properties of the 14 spectroscopically confirmed luminous LAEs at $z\approx5.7$ based on the JWST imaging data in the four NIRCam bands (F115W, F150W, F277W, and F444W).

The paper has a layout as follows. In Section 2, we briefly describe the sample of luminous LAEs at $z\approx5.7$, COSMOS-Web NIRCam imaging observations, data reduction, and photometry. We present the results of a variety of properties about the LAE sample in Section 3. We give further discussions in Section 4. We summarize our paper in Section 5. Throughout the paper, we use a standard flat cosmology with $H_0=\rm{70\ km\ s^{-1}\ Mpc^{-1}}$, $\Omega_m=0.3$ and $\Omega_{\Lambda}=0.7$. All magnitudes refer to the AB system.

\begin{figure*}[t]
\centering
\gridline{\fig{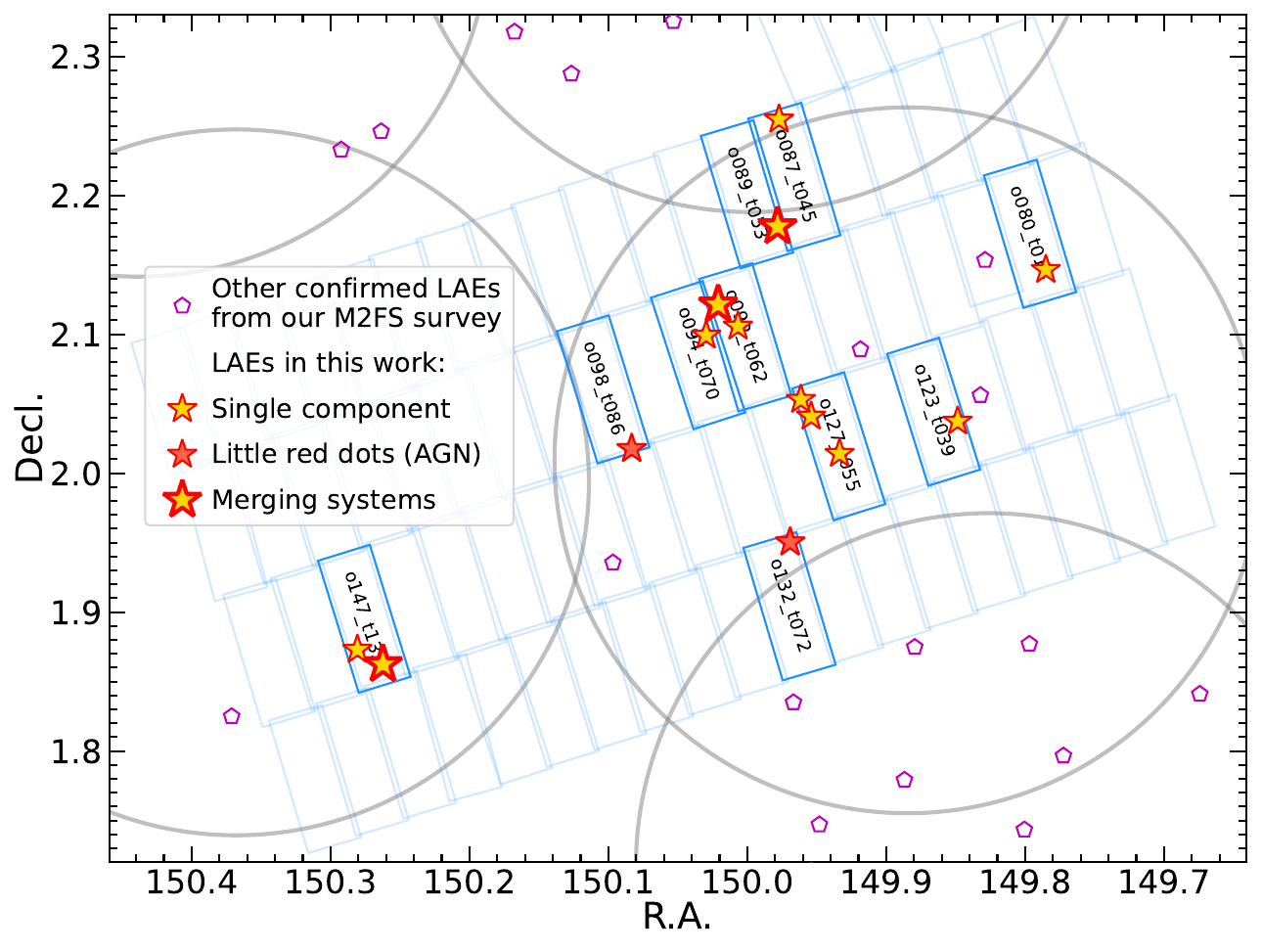}{0.69\textwidth}{(a)}
\fig{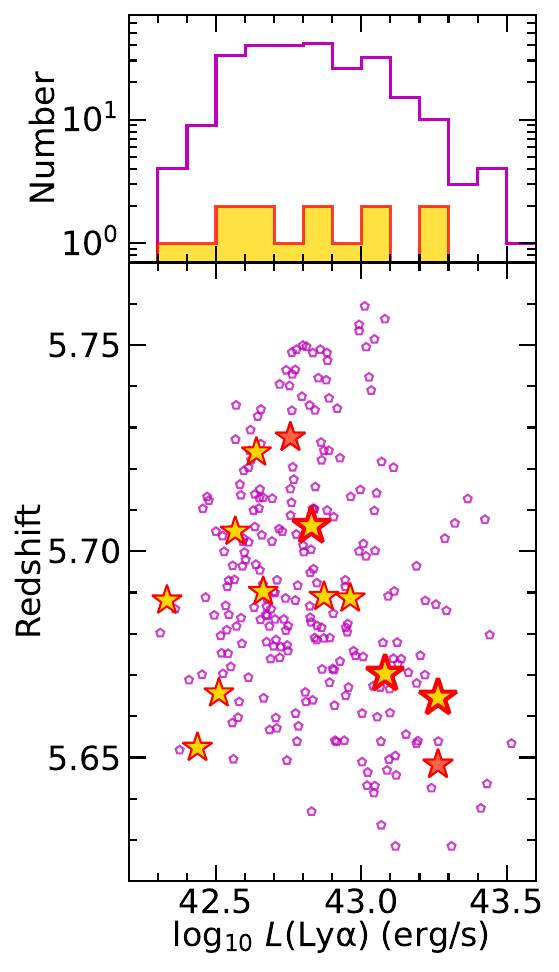}{0.3\textwidth}{(b)}}
\gridline{\fig{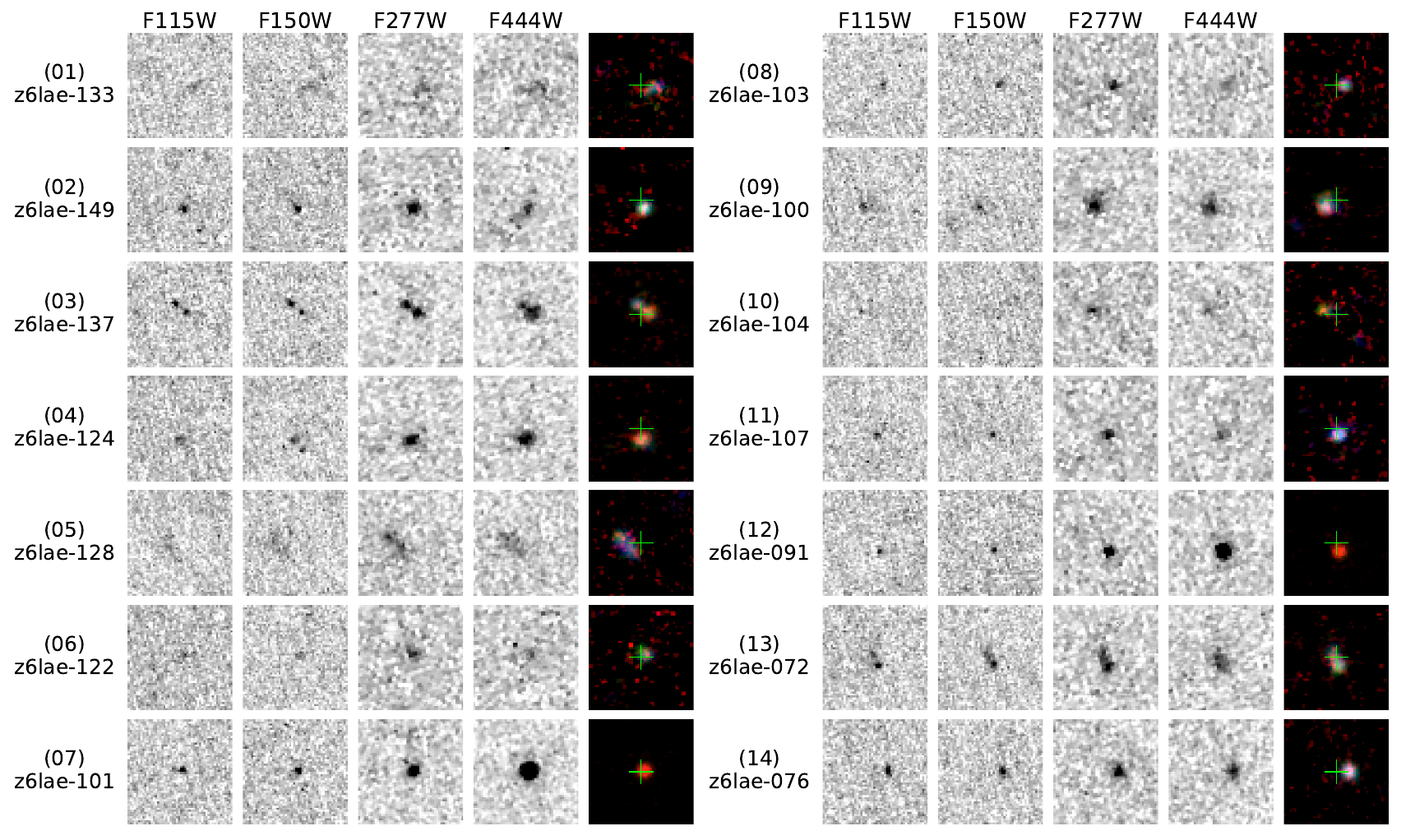}{0.95\textwidth}{(c)}}
\caption{\textbf{(a)} Our LAE sample covered by the COSMOS-Web observations. The stars represent the 14 LAEs, in which the larger ones with thicken edges indicate the merging/interacting systems (Section \ref{sec:pair}) while the red-filled ones indicate the little red dots with AGN features (Section \ref{sec:agn}). The magenta pentagons correspond to all other $z\approx5.7$ LAEs spectroscopically confirmed by our M2FS spectroscopic survey. The visit tiles covering our sample have a darker edge with the observation IDs (Column 5 in Table \ref{tab1}). The big circles represent the M2FS pointings. 
\textbf{(b)} Lower panel gives the redshift--\lya\ luminosity
distribution of the LAEs. The symbols are same as those in the lefthand Figure~\ref{samp}(a). Note that two LAEs, z6lae-072 and z6lae-076, overlap each other in the parameter space. The corresponding number distributions are plotted shown in upper panel.
\textbf{(c)} The 14 LAEs detected in the four JWST/NIRCam bands ($1\arcsec.5\times1\arcsec.5$, north is up and east to left) with pseudo-color images (after PSF matching; blue: F115W+F150W, green: F277W, red: F444W). The central crosses mark the \lya\ centroids detected from the NB816 images.
\label{samp}
}
\end{figure*}

\section{Sample \& Data}

In this section, we describe our sample of LAEs at $z\approx5.7$, \jwst/NIRCam imaging observations of the COSMOS-Web program, imaging data reduction, and photometry. We list the observational information of the sample in Table~\ref{tab1}.

\subsection{Sample of LAEs at $z\approx5.7$}

To build a large and homogeneous sample of \hz\ galaxies (LBGs at $z\sim6$ and LAEs at $z\approx5.7$ and 6.6), we have carried out a spectroscopic survey with a total sky coverage of ${\sim}2$ deg$^2$. As designed by our survey, the observed galaxy candidates are selected from Subaru/Suprime broadband and narrowband (NB) images in several well-studied fields including the Subaru {\it XMM-Newton} Deep Survey (SXDS), the Extended {\it Chandra} Deep Field-South (ECDFS), A370, COSMOS, and SSA22. The utilized instrument in this survey is the fiber-fed, multi-object spectrograph Michigan/Magellan Fiber System (M2FS; \citealt{mateo12}) on the 6.5 m Magellan Clay telescope. In the five fields, we have identified 260 LAEs at $z\approx5.7$ which by far construct the largest spectroscopically confirmed sample of LAEs at this redshift \citep{ning20}. In COSMOS, we confirm 52 LAEs at $z\approx5.7$ from 158 observed candidates. In this work, we focus on 14 LAEs therein which have been covered by the COSMOS-Web JWST/NIRCam imaging observations.

We show the positions of our confirmed LAEs embedded within the COSMOS-Web map of tiling visits in Figure~\ref{samp}a. Among the 18 LAEs covered by COSMOS-Web observations, we do not include four LAEs in the sample. The reason is that three of them (z6lae-089, z6lae-110 and z6lae-119) are not detected in both F115W and F150W images (before the PSF matching to F444W; see Section \ref{sec:dophot}) while the another one is located in the tile but with unfinished observations. So we have 14 LAEs in our final sample. Note that in the lefthand region, there are only two LAEs from our sample. This is because two M2FS pointings, COSMOS1 and COSMOS3, suffered serious alignment problems as mentioned in \citet{ning20}. The number density of confirmed LAEs is relatively low in the regions. In Figure~\ref{samp}b, we plot how the 14 LAEs distribute in the redshift--luminosity space in the lower panel and the number distribution in the upper panel. We can see that the 14 LAEs at $z\approx5.7$ cover a range of \lya\ luminosity over an order of magnitude, which indicate that our sample is representative in term of \lya\ luminosity to study properties of LAEs at this redshift slice.

\subsection{COSMOS-Web Imaging Data}
Our LAE sample have been covered by the ongoing \jwst\ Cycle 1 program COSMOS-Web \citep[GO 1727;][]{cosmosWebb, casey23} with quick public data release. COSMOS-Web survey aims to observe a contiguous central area of 0.54 deg$^2$ in the COSMOS field. It carries out NIRCam imaging observations in four broad bands including F115W and F150W in the short wavelength (SW), and F277W and F444W in the long wavelength (LW). COSMOS-Web also has a parallel MIRI imaging in the F770W band, which cover a non-contiguous, smaller area of 0.19 deg$^2$. The whole COSMOS-Web program own 152
separate visits arranged in a $19\times8$ grid. After the first half of observations, the 14 LAEs in our sample locates in 10 mosaic tiles (Figure~\ref{samp}a) whose exposure time in all the four bands have reached the designed time of 1031s. We retrieve the uncalibrated NIRCam imaging data from the Mikulski Archive for Space Telescopes (MAST) Archive at the Space Telescope Science Institute. The specific observations can be accessed via\dataset[DOI: 10.17909/vvwq-8284]{https://doi.org/10.17909/vvwq-8284}.

\subsection{Data Reduction and Photometry}
\label{sec:dophot}

We reduce the NIRCam imaging data with the standard \jwst\ pipeline\dataset[v1.7.2]{https://zenodo.org/records/7071140} \citep{jwst_v1p7} up to stage 2 using the reference files “jwst$\_$1046.pmap”. Then we use the \texttt{Grizli}\footnote{\url{https://github.com/gbrammer/grizli}} reduction pipeline to process the output images. \texttt{Grizli} mitigates 1/f noises and mask the “snowball” artifacts from cosmic rays \citep{rigby23}.  It further converts the world coordinate system (WCS) information in the headers to the SIP format for each exposure so that images can be drizzled and combined with Astrodrizzle\footnote{\url{https://drizzlepac.readthedocs.io/en/latest/astrodrizzle.html}}. For the SW and LW images, the WCS of final mosaics are registered based on the catalogs of DESI Legacy Imaging Surveys Data Release 9 and the pixel scale is resampled to 0\arcsec.03 with pixfrac $=0.8$. We also subtract an additional background on the final mosaics. Figure~\ref{samp}c shows the thumbnail images for our sample of LAEs at $z\approx5.7$ in the four JWST/NIRCam bands.

We run \texttt{SExtractor} \citep{ber96} to detect the sources and measure the flux in the \jwst/NIRCam four band images. As the original target positions are given by the narrow-band (NB) data (\lya\ centroid), we first match the output catalogs to the targets within a distance tolerance of $0\arcsec.3$. For each target, we select its highest-S/N band to feed the detection image. Usually, we adopt the F277W or F444W (boosted by \ha\ emission) band for most of sources otherwise we adopt the SW bands for the multiple components in the merging/interacting systems, z6lae-072 and z6lae-137. We then rerun \texttt{SExtractor} in the dual image mode to perform photometry with the corresponding detection images. In this procedure, we create an empirical point-spread function (PSF) by co-adding at least 100 bright (not saturated) stars for each band and carry out PSF matching to the F444W band. Then we adopting an aperture with a radius of 0\arcsec.15 in each measurement image. The aperture correction is calculated from the empirical PSF in the F444W images. For those containing double components, we adopt the original image for each band in case of source blending and do photometry using an aperture with a diameter of the distance between the two components. We also improve the photometric measurements by performing \texttt{GALFIT} (see Section \ref{sec:galmorp}). We perform simulations to evaluate the flux uncertainty. For each galaxy (detected in the band), we create 100 mock sources (from \texttt{GALFIT}), and put them randomly (around the galaxy) on the given image. We then use \texttt{SExtractor} as we did photometry for the real objects, and compute the mean and standard deviation of the measurements. Table \ref{tab1} lists the multi-band photometry results of the galaxy sample.

\floattable
\renewcommand{\arraystretch}{1.0}
\begin{deluxetable}{lccccccccc}[t]
\tablecaption{Observational information of the COSMOS-Web covered LAEs at $z\approx5.7$.
\label{tab1}}
\centering
\tablehead{
   \colhead{ID} & \colhead{R.A.} & \colhead{Decl.} & \colhead{$z_{Ly\alpha}$} & \colhead{ObsID} & \colhead{NB816} & \colhead{F115W} & \colhead{F150W} & \colhead{F277W} & \colhead{F444W} \\
   \colhead{} & \colhead{(J2000.0)} & \colhead{(J2000.0)} & \colhead{} & \colhead{} & \colhead{(mag)} & \colhead{(mag)} & \colhead{(mag)} & \colhead{(mag)} & \colhead{(mag)} \\
   \colhead{(1)} & \colhead{(2)} & \colhead{(3)} & \colhead{(4)} & \colhead{(5)} & \colhead{(6)} & \colhead{(7)} & \colhead{(8)} & \colhead{(9)} & \colhead{(10)}
   }
\startdata
z6lae-133 & 09:59:08.43 & $+$02:08:47.3 & 5.688 & o080-t014 & 25.86$\pm$0.11 & 27.67$\pm$0.27 & 27.74$\pm$0.23 & 27.72$\pm$0.12 & 27.45$\pm$0.09 \\
z6lae-149 & 09:59:54.52 & $+$02:15:16.6 & 5.689 & o087-t045 & 24.66$\pm$0.03 & 26.51$\pm$0.13 & 26.89$\pm$0.15 & 26.60$\pm$0.05 & 26.62$\pm$0.06 \\
z6lae-137 & 09:59:54.78 & $+$02:10:39.3 & 5.665 & o089-t053 & 24.31$\pm$0.02 & 26.56$\pm$0.08 & 26.68$\pm$0.07 & 26.41$\pm$0.04 & 26.13$\pm$0.04 \\
z6lae-137a & 09:59:54.77 & $+$02:10:39.3 & 5.665 & o089-t053 & ... & 27.44$\pm$0.13 & 27.83$\pm$0.15 & 27.04$\pm$0.05 & 26.60$\pm$0.04 \\
z6lae-137b & 09:59:54.78 & $+$02:10:39.4 & 5.665 & o089-t053 & ... & 27.20$\pm$0.10 & 27.14$\pm$0.08 & 27.30$\pm$0.06 & 27.26$\pm$0.07 \\
z6lae-124 & 10:00:01.61 & $+$02:06:20.4 & 5.705 & o092-t062 & 25.31$\pm$0.06 & 27.20$\pm$0.25 & 26.91$\pm$0.16 & 26.55$\pm$0.05 & 25.99$\pm$0.04 \\
z6lae-128 & 10:00:05.04 & $+$02:07:17.1 & 5.706 & o092-t062 & 24.96$\pm$0.04 & 27.02$\pm$0.15 & 26.94$\pm$0.11 & 27.03$\pm$0.06 & 26.95$\pm$0.07 \\
z6lae-128a & 10:00:05.05 & $+$02:07:17.0 & 5.706 & o092-t062 & ... & 27.35$\pm$0.15 & 27.87$\pm$0.18 & 27.87$\pm$0.09 & 27.65$\pm$0.09 \\
z6lae-128b & 10:00:05.06 & $+$02:07:17.1 & 5.706 & o092-t062 & ... & $>$28.79 & 27.54$\pm$0.13 & 27.71$\pm$0.08 & 27.77$\pm$0.10 \\
z6lae-122 & 10:00:07.12 & $+$02:05:57.0 & 5.666 & o094-t070 & 25.70$\pm$0.08 & 27.54$\pm$0.32 & 27.87$\pm$0.34 & 27.25$\pm$0.08 & 27.48$\pm$0.11 \\
z6lae-101 & 10:00:19.97 & $+$02:01:03.3 & 5.648 & o098-t086 & 24.80$\pm$0.04 & 26.75$\pm$0.17 & 26.54$\pm$0.11 & 26.00$\pm$0.04 & 24.34$\pm$0.01 \\
z6lae-103 & 09:59:23.66 & $+$02:02:14.4 & 5.724 & o123-t039 & 25.53$\pm$0.06 & 27.22$\pm$0.23 & 27.64$\pm$0.27 & 27.52$\pm$0.11 & 27.14$\pm$0.09 \\
z6lae-100 & 09:59:44.06 & $+$02:00:50.7 & 5.689 & o127-t055 & 24.63$\pm$0.03 & 26.20$\pm$0.12 & 26.42$\pm$0.12 & 26.32$\pm$0.05 & 26.11$\pm$0.04 \\
z6lae-104 & 09:59:48.97 & $+$02:02:27.1 & 5.652 & o127-t055 & 25.97$\pm$0.08 & 28.11$\pm$0.49 & $>$28.53 & 27.77$\pm$0.11 & 27.70$\pm$0.11 \\
z6lae-107 & 09:59:50.75 & $+$02:03:10.7 & 5.690 & o127-t055 & 25.28$\pm$0.05 & 26.50$\pm$0.15 & 26.96$\pm$0.19 & 27.14$\pm$0.07 & 27.12$\pm$0.08 \\
z6lae-091 & 09:59:52.60 & $+$01:57:01.1 & 5.728 & o132-t072 & 25.32$\pm$0.05 & 27.94$\pm$0.46 & 27.61$\pm$0.28 & 26.62$\pm$0.05 & 25.01$\pm$0.02 \\
z6lae-072 & 10:01:02.95 & $+$01:51:44.8 & 5.670 & o147-t135 & 24.54$\pm$0.03 & 26.36$\pm$0.07 & 26.37$\pm$0.06 & 26.26$\pm$0.04 & 26.09$\pm$0.05 \\
z6lae-072a & 10:01:02.95 & $+$01:51:44.9 & 5.670 & o147-t135 & ... & 27.43$\pm$0.13 & 27.31$\pm$0.09 & 27.21$\pm$0.07 & 26.92$\pm$0.07 \\
z6lae-072b & 10:01:02.95 & $+$01:51:44.7 & 5.670 & o147-t135 & ... & 26.87$\pm$0.08 & 26.97$\pm$0.07 & 26.84$\pm$0.05 & 26.77$\pm$0.06 \\
z6lae-076 & 10:01:07.36 & $+$01:52:22.7 & 5.670 & o147-t135 & 24.59$\pm$0.03 & 26.78$\pm$0.13 & 26.79$\pm$0.10 & 26.86$\pm$0.06 & 26.71$\pm$0.05 
\enddata
\centering
\tablecomments{The target coordinates are from the original LAE catalog given by NB detections \citep{ning20} except the components of the merging systems with the suffixes $a$ and $b$ (the NB magnitudes are not shown) given by the NIRCam detections. The greater-than symbols indicate $2\sigma$ upper limits.}
\end{deluxetable}

\section{Results}

In this section, we present the results of a variety of galaxy properties. We first study the galaxy morphology and then constrain the rest-frame UV continua. Next we perform SED fitting to constrain their stellar mass ($M_*$) and star-formation rates (SFR). We also utilize previous \lya\ data to investigate the misalignment between the \lya\ and UV emission. The results are summarized in Table \ref{tab2}.

\floattable
\renewcommand{\arraystretch}{1.}
\begin{deluxetable*}{lcccccccc}[t]
\tablecaption{Measured properties of the LAE sample at redshift 5.7.
\label{tab2}}
\centering
\tablehead{
   \colhead{ID} & \colhead{${\rm log}_{10} L$(\lya)} & \colhead{\ewlya} & \colhead{\Dd} & \colhead{$R_{\rm e}$} & \colhead{$M_{\rm UV}$} & \colhead{$\beta_{\rm UV}$} & \colhead{${\rm log}_{10} M_*$} & \colhead{SFR} \\
   \colhead{} & \colhead{($\rm erg\ s^{-1}$)} & \colhead{(\AA)} & \colhead{(kpc)} & \colhead{(kpc)} & \colhead{} & \colhead{} & \colhead{($M_{\odot}$)} & \colhead{($M_{\odot}$ yr$^{-1}$)} \\
   \colhead{(1)} & \colhead{(2)} & \colhead{(3)} & \colhead{(4)} & \colhead{(5)} & \colhead{(6)} & \colhead{(7)} & \colhead{(8)} & \colhead{(9)}
   }
\startdata
z6lae-133 & 42.45 $\pm$ 0.17 & 125 $\pm$ 50  & 1.27 $\pm$ 0.60 & 0.47 $\pm$ 0.10 & -18.98 $\pm$ 0.42 & -2.26 $\pm$ 1.25 & $8.56_{-0.24}^{+0.15}$ & $1.34_{-0.33}^{+0.69}$ \\
z6lae-149 & 42.90 $\pm$ 0.09 & 107 $\pm$ 21  & 0.62 $\pm$ 0.28 & 0.18 $\pm$ 0.03 & -20.28 $\pm$ 0.21 & -3.31 $\pm$ 0.70 & $8.69_{-0.22}^{+0.17}$ & $3.71_{-0.68}^{+1.01}$ \\
z6lae-137 & 43.24 $\pm$ 0.05 & 272 $\pm$ 33  & 0.80 $\pm$ 0.23 & ... & -20.10 $\pm$ 0.13 & ... & ... & ... \\
z6lae-137a & ... & ... & ... & $<$0.21 & -19.36 $\pm$ 0.21 & -3.38 $\pm$ 0.69 & $9.12_{-0.09}^{+0.07}$ & $3.06_{-0.59}^{+1.01}$ \\
z6lae-137b & ... & ... & ... & 0.13 $\pm$ 0.04 & -19.37 $\pm$ 0.16 & -1.77 $\pm$ 0.47 & $8.25_{-0.21}^{+0.16}$ & $2.75_{-0.65}^{+1.18}$ \\
z6lae-124 & 42.69 $\pm$ 0.15 & 166 $\pm$ 59  & 0.84 $\pm$ 0.41 & 0.28 $\pm$ 0.04 & -19.28 $\pm$ 0.38 & -1.00 $\pm$ 1.04 & $9.27_{-0.19}^{+0.13}$ & $8.72_{-2.60}^{+2.90}$ \\
z6lae-128 & 42.83 $\pm$ 0.09 & 177 $\pm$ 39  & 1.11 $\pm$ 0.33 & ... & -19.55 $\pm$ 0.23 & ... & ... & ... \\
z6lae-128a & ... & ... & ... & 0.22 $\pm$ 0.12 & -19.51 $\pm$ 0.24 & -3.80 $\pm$ 0.82 & $8.18_{-0.20}^{+0.20}$ & $1.47_{-0.32}^{+0.44}$ \\
z6lae-128b & ... & ... & ... & 0.65 $\pm$ 0.06 & $>$-18.83 & ... & ... & ... \\
z6lae-122 & 42.61 $\pm$ 0.21 & 143 $\pm$ 67  & 0.60 $\pm$ 0.54 & 0.23 $\pm$ 0.08 & -19.23 $\pm$ 0.51 & -3.17 $\pm$ 1.63 & $8.68_{-0.30}^{+0.14}$ & $3.15_{-1.80}^{+4.89}$ \\
z6lae-101 & 43.29 $\pm$ 0.11 & 428 $\pm$ 106 & 0.25 $\pm$ 0.30 & 0.10 $\pm$ 0.03 & -19.74 $\pm$ 0.27 & -1.27 $\pm$ 0.71 & ... & ... \\
z6lae-103 & 42.62 $\pm$ 0.15 & 104 $\pm$ 36  & 0.54 $\pm$ 0.47 & 0.18 $\pm$ 0.05 & -19.60 $\pm$ 0.37 & -3.48 $\pm$ 1.25 & $8.65_{-0.23}^{+0.17}$ & $1.84_{-0.51}^{+0.94}$ \\
z6lae-100 & 42.91 $\pm$ 0.08 & 86 $\pm$ 16  & 1.06 $\pm$ 0.27 & 0.51 $\pm$ 0.08 & -20.52 $\pm$ 0.19 & -2.79 $\pm$ 0.61 & $8.99_{-0.13}^{+0.11}$ & $4.91_{-0.93}^{+1.32}$ \\
z6lae-104 & 42.70 $\pm$ 0.30 & 288 $\pm$ 197 & 1.31 $\pm$ 0.64 & 0.22 $\pm$ 0.14 & -18.70 $\pm$ 0.74 & ... & ... & ... \\
z6lae-107 & 42.54 $\pm$ 0.10 & 44 $\pm$ 10  & 0.39 $\pm$ 0.40 & 0.22 $\pm$ 0.05 & -20.35 $\pm$ 0.25 & -3.64 $\pm$ 0.85 & $8.25_{-0.21}^{+0.18}$ & $3.05_{-0.61}^{+0.80}$ \\
z6lae-091 & 42.77 $\pm$ 0.28 & 393 $\pm$ 251 & 0.61 $\pm$ 0.41 & 0.10 $\pm$ 0.05 & -18.53 $\pm$ 0.69 & -0.87 $\pm$ 1.89 & ... & ... \\
z6lae-072 & 43.07 $\pm$ 0.05 & 161 $\pm$ 17  & 0.21 $\pm$ 0.26 & ... & -20.25 $\pm$ 0.11 & ... & ... & ... \\
z6lae-072a & ... & ... & ... & 0.13 $\pm$ 0.03 & -19.11 $\pm$ 0.20 & -1.57 $\pm$ 0.55 & $8.74_{-0.20}^{+0.12}$ & $2.31_{-0.63}^{+1.23}$ \\
z6lae-072b & ... & ... & ... & 0.43 $\pm$ 0.05 & -19.78 $\pm$ 0.13 & -2.35 $\pm$ 0.38 & $8.70_{-0.18}^{+0.11}$ & $2.87_{-0.46}^{+0.91}$ \\
z6lae-076 & 43.07 $\pm$ 0.08 & 239 $\pm$ 45  & 1.01 $\pm$ 0.27 & 0.22 $\pm$ 0.03 & -19.83 $\pm$ 0.20 & -2.03 $\pm$ 0.58 & $8.65_{-0.16}^{+0.13}$ & $3.17_{-0.55}^{+0.84}$ 
\enddata
\centering
\tablecomments{The less-than sign indicates $2\sigma$ upper limits.}
\end{deluxetable*}

\subsection{Target Appearance and Morphology}
\label{sec:galmorp}

As shown in Figure~\ref{samp}c, most of LAEs in our sample are either compact or irregular in the JWST/NIRCam images.  Besides them, three LAEs (z6lae-072, z6lae-128, and z6lae-137) are resolved with interacting or double-component features. With \lya\ redshifts, we obtain their projected distances of 0\arcsec.16 (0.93), 0\arcsec.23 (1.32), and 0\arcsec.19 (1.12 kpc), respectively. It is almost impossible that one of the components is a foreground or background object due to a very low probability (${\lesssim}1\%$) that two random sources are located within an angular distance of $0\arcsec.2$ \citep{fusq23}. Among the three LAEs, only z6lae-137 is covered by HST/ACS F606W imaging observations. No detection ($3\sigma$ upper limit of 27.8 mag in an aperture with a radius of 0\arcsec.15) exists at the positions of its two components, which supports they are both at $z>5$. Note that we cannot confirm robust (major or minor) mergers without the velocity-separation information from rest-frame optical spectra \citep[e.g.,][]{ventou17, daiy21}. As the luminous LAEs are usually small with a typical size of ${\lesssim}0.5$ kpc (see our following result of galaxy sizes), the components do not tend to represent star-forming clumps within the same galaxy. They are thus probably in an advanced merging stage due to their close proximity considering more frequent merger events at high redshift \citep[e.g.,][]{RodriguezGomez15}. In this work, we also refer to them using the term ``pair" for brevity.

In this work, we measure galaxy size to quantify the morphology of the LAEs (and their components). We use the half-light radius \Reff, the radius containing half of the total light, to represent the galaxy size. We measure the galaxy size at the rest-frame UV band which fall in the NIRCam SW filters with higher resolution than the LW filters. Before the measurement, we combine both F115W (with the PSF matched to F150W) and F150W bands to create stacking rest-UV images with higher signal-to-noise ratio (S/N). As F115W and F150W bands are next to each other in the wavelength range, we ignore the effect of the morphological $k$-correction. The stacked images thus correspond to the rest-frame UV continuum without the influence from strong nebular emission line for our galaxy sample. We run \texttt{GALFIT} \citep{peng02, peng10} to directly measure the intrinsic effective radius by modeling a Sérsic profile. For each LAE covered by different visit tile of COSMOS-Web, we adopt an empirical PSF by co-adding at least 20 bright (not saturated) stars in each F150W tile image. The measured \Reff\ (along the semi-major axis) is then circularized to represent the galaxy size by multiplying the square root of the corresponding axis ratio. The \texttt{GALFIT} results are listed in the Column 5 of Table \ref{tab2}. Note that no size is shown for z6lae-104 due to its faintness in both F115W and F150W.

\begin{figure}[t]
\epsscale{1.15}
\centering
\plotone{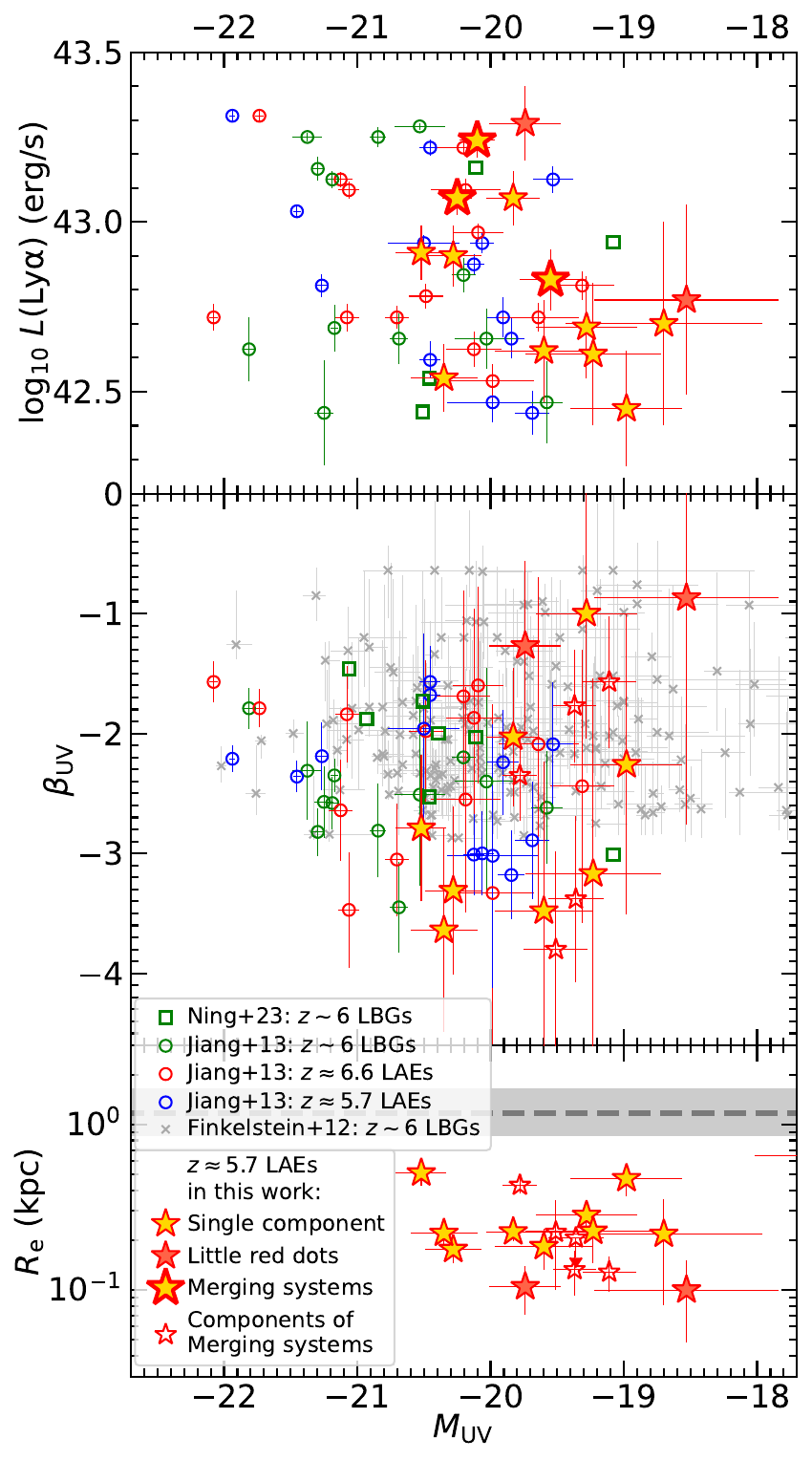}
\caption{\lya\ luminosity (upper panel), UV slope (middle panel), and half-light radius (lower panel) as a function of the absolute UV magnitude at rest-frame 1500 \AA. The star symbols indicate the 14 LAEs $z\approx5.7$ in this work. Note that the larger stars with thicken edges mark the merging/interacting systems and the smaller open ones mark their individual components. The green squares represent the LBGs (luminous in \lya) at $z\simeq6$ from \citet{ning22}. The open circles correspond to the LBGs at $z\simeq6$ (green), LAEs at $z\approx5.7$ (blue) and 6.5 (red) from \citet{jiang13a}. The gray crosses represent the LBGs at $z\sim6$ from \citet{finkelstein12}. In the lower panel, the gray dashed line with an uncertainty region indicates an average \Reff\ for galaxies at $z\approx5.7$, which is obtained from the size-evolution results of \citet{ormerod24}.
\label{uv}}
\end{figure}

The intrinsic half-light radius \Reff\ is plotted as a function of \muv\ in the lower panel of Figure~\ref{uv}. All of the LAEs lie below the average value of \Reff\ shown by the gray shaded regions. Note that among the sample, the two reddest LAEs in F277W$-$F444W (z6lae-091 and z6lae-101) have smallest sizes close to PSF. For the rest, \Reff\ are between 0.1 kpc and 0.7 kpc with a median \Reff\ of ${\approx}0.2$ kpc. Such statistics show broad consistency with \citet{jiang13b} which obtained a median \Reff\ of ${\sim}0.9$ kpc because our UV-fainter sample naturally appear smaller considering the size-luminosity relation. The individual galaxy in merging/interacting systems span a similar \Reff\ range. Our results are also broadly consistent with other previous works about \hz\ LAEs based on \hst\ imaging data \citep[e.g.,][]{taniguchi09, Paulino-Afonso18}. The size-luminosity relation is not apparent for our sample whose number of sources is currently limited.

\subsection{Rest-frame UV Continua}
\label{sec:uv}

Thanks to the deep near-IR photometry based on the \jwst\ facilities, we can constrain weak UV continua of the faint galaxies. We measure the galaxy UV continua following our previous work \citep{ning22}. In this measurement, two objects, z6lae-104 and z6lae-128b, are discarded due to the ${<}2\sigma$ detection(s) in F115W or F150W bands.
Within the rest of sample, only one source (z6lae-122) does not have ${>}4\sigma$ detections in both F115W and F150W. For all other sources with ${>}4\sigma$ detections in at least one SW band, the other band still has ${>}3\sigma$ detections except z6lae-091.

We assume a power-law form for the UV continuum of each source, i.e.: $f_\lambda \propto \lambda^{\beta}$. In AB magnitude units, it has a linear relation $m_{\rm AB} \propto (\beta+2)\times{\rm log}(\lambda)$. We then constrain the UV continuum using the SW photometric data for each source. We thus estimate the UV-continuum slope \buv\ with a single UV color. For our LAE sample at $z\approx5.7$, the F115W and F150W do not cover \lya\ emission. We next subtract the UV continuum from the NB816 photometry to re-constrain \lya\ flux (luminosity) and estimate the \lya\ equivalent width \ewlya. The intergalactic medium (IGM) absorption blueward of \lya\ line is considered in the computation based on the model from \citet{madau95}. The magnitude errors propagate into the measured UV quantities.

We plot the \lya\ luminosity as a function of the absolute UV magnitude in Figure~\ref{uv} (upper panel). The measured \muv\ of the sample spread between ${-}20.5$ and ${-}18.5$ with a median value of $-19.7$. The \lya\ luminosity has a potential dependence on \muv. Note that none of the LAEs have \muv~$<-20.6$ in our sample. The sample also shifts towards fainter UV luminosities when comparing to other samples. The reason is that \citet{jiang13a} exclude galaxies whose broadband photometry has weak detections of ${<}5\sigma$ in the $J$ band or ${<}3\sigma$ in any other band.

In the middle panel of Figure~\ref{uv}, we show the relation between \buv\ and \muv. Our sample have a broad \buv\ range roughly from ${-}3.5$ to ${-}1.0$. The median value is \buv~$=-2.35\pm0.42$ which is broadly consistent with previous works \citep[e.g.,][]{finkelstein12, jiang13a, bouwens14, simmonds23}. But the luminous LAEs in our sample are overall bluer. Among them, six sources including two components in the merging/interacting systems have possibly extreme UV slopes of \buv~$<-3$ with a mean $-3.46\pm0.43$. The extreme blue slopes usually indicate the existence of very young populations with low metallicity \citep[e.g.,][]{jiang20a, topping22}. It is worth noting that for z6lae-137, which is a close pair, the two components have very different UV slopes. Their internal properties are thus supposed to be largely diverse, such as the dust attenuation, metallicity and age of the stellar populations, which also supports that they are two individual galaxies in the merging system (in comparison with the sample from \citealt{bolamperti23}). The above results need to be verified by larger samples with more accurate UV-slope measurements from more JWST SW bands.

\begin{figure*}
\centering
\includegraphics[angle=0, width=0.9\textwidth]{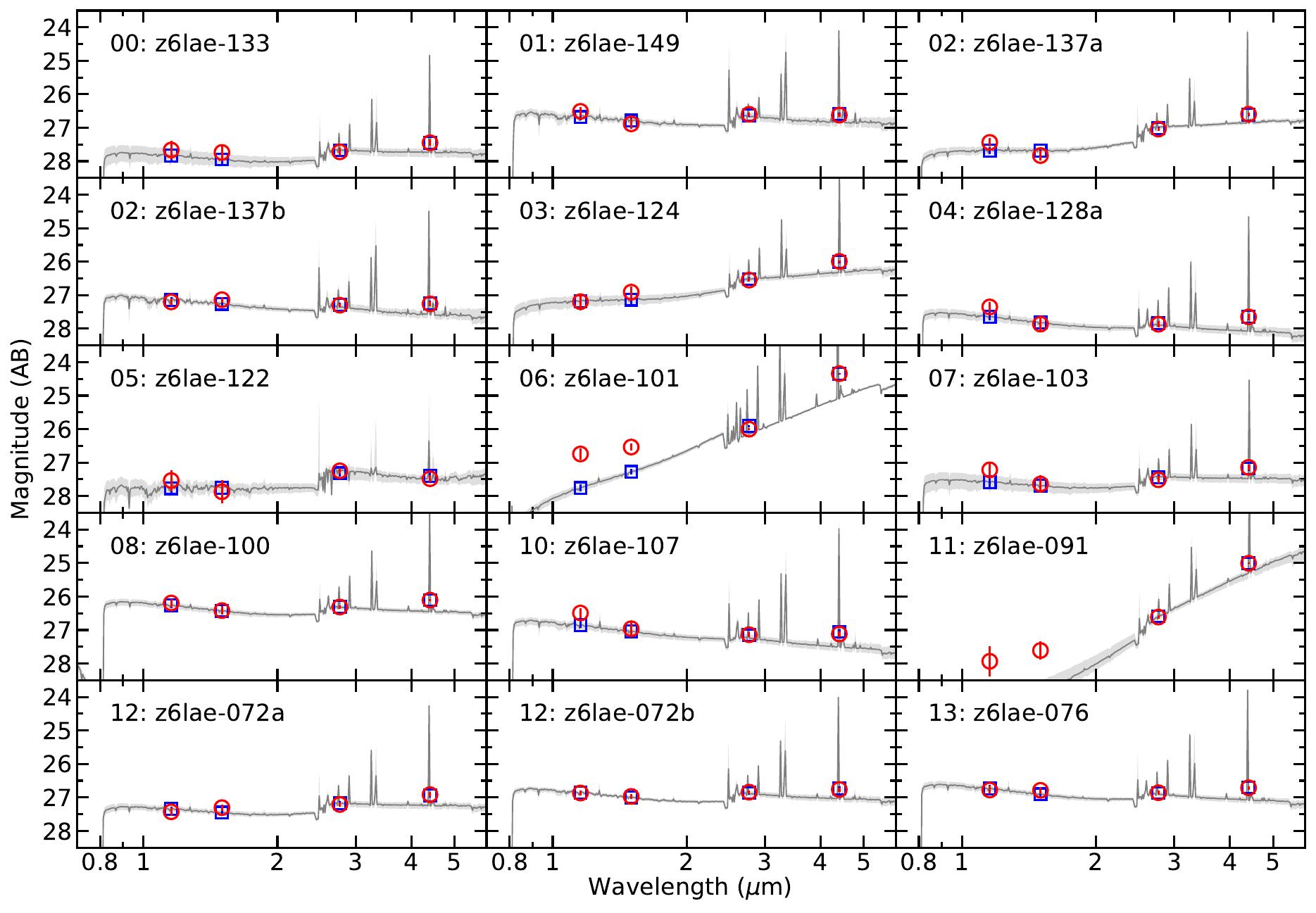}
\caption{SED fitting for our LAE sample with \texttt{Bagpipes}. In each panel, the observed and best-fit photometric data are shown by red circles and blue squares with error bars, respectively. The best-fit model SED is plotted as a gray line with a shaded region indicating $1\sigma$ uncertainty. Note that the best-fit model SEDs can not match the observed SEDs for the two little red dots, z6lae-091 and z6lae-101, which imply probably additional AGN components.
\label{sed}}
\end{figure*}

\begin{figure*}
\centering
\includegraphics[angle=0, width=0.85\textwidth]{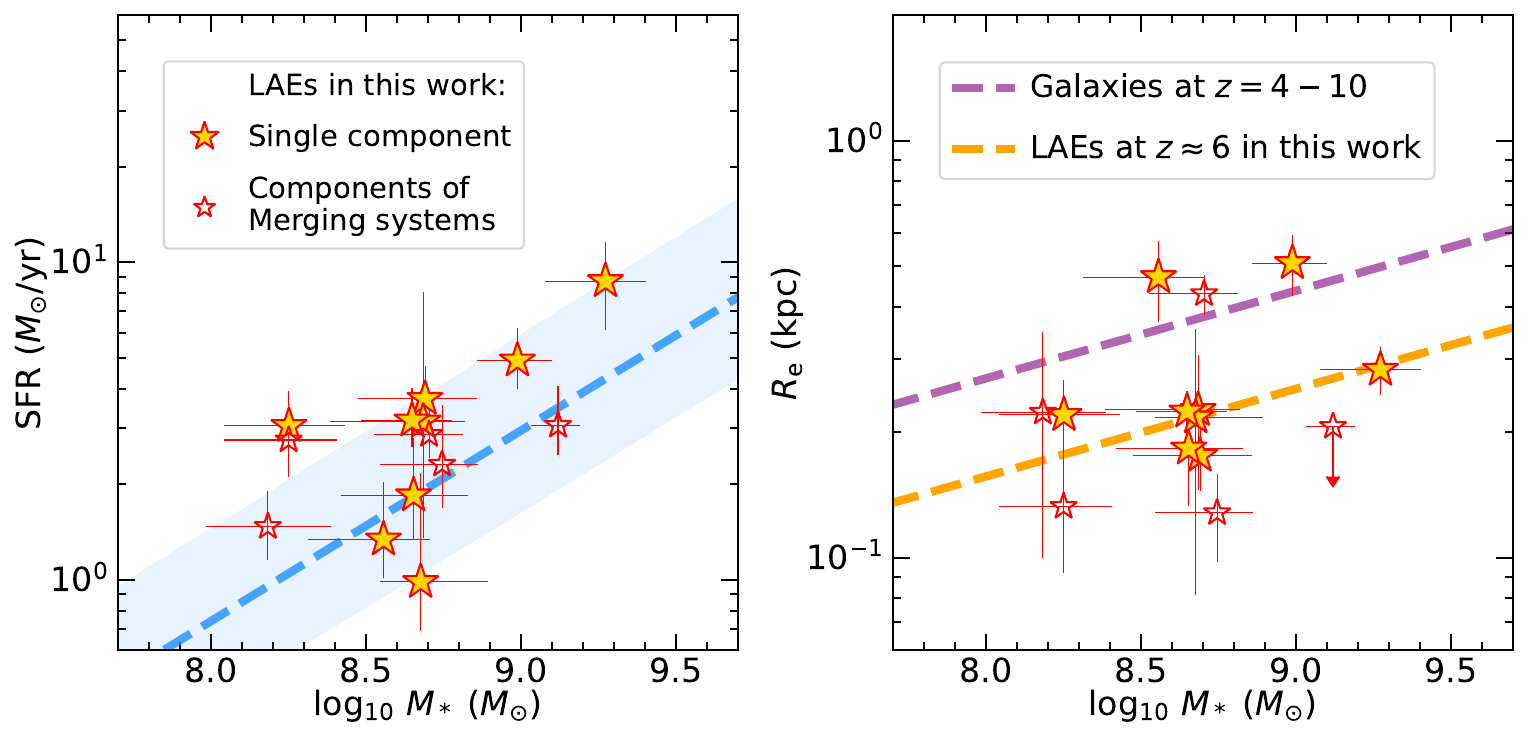}
\caption{The SFR-$M_*$ and size-$M_*$ distribution of our LAE sample. The stars indicate our sample of 14 LAEs at $z\approx5.7$, in which the smaller open ones mark the components of the merging/interacting systems. In the left panel, the blue lines represent the best-fit MS power-law relation from \citet{rinaldi22}. In the right panel, the orange dashed line is the best-fit linear relation for our LAEs at $z\approx5.7$. The purple line gives the mass-size relation for galaxies at $z=4-10$ from \citet{2023arXiv230706336L}.
\label{pps}}
\end{figure*}

\subsection{Physical Properties}
\label{sec:pps}

We constrain the physical properties of the 14 LAEs at $z\approx5.7$ by adopting SED fitting. These LAEs have spectroscopic redshifts from \lya\ lines. The spectroscopic redshift is a significant parameter when performing SED fitting. Without the secure redshift, the nebular lines of galaxies may be set into wrong filter bands. We adopt the SED-fitting tool \texttt{Bagpipes} \citep{carnall18}. \texttt{Bagpipes} provides a Python framework for self-consistently modeling the stellar, nebular, dust, and IGM properties of galaxy spectra.

A small fraction of our sample are very blue with likely extreme slopes of ${\lesssim}{-3}$. \citet{jiang20a} studied six very blue LAEs at $z\simeq6$. They found that young populations without dust content can produce very blue slopes reaching $\beta\simeq-3$, but the SED models can not produce extreme slopes of $\beta<-3$. This work used three or four photometric data points obtain relatively reliable UV slopes. However, we currently compute the UV slopes based on the two SW bands. As the UV slopes are not enough robust with large errors due to only two (low-S/N) SW photometric data, we keep adopting a standard SED fitting procedure based on \texttt{Bagpipes} even for the bluest LAEs in this work. The results show that the best-fit SED models agree with the observational photometry within the uncertainty range although they can not perfectly match the UV colors. Two galaxies (z6lae-128b and z6lae-104) in our sample are not detected (${<}3\sigma$) in F115W or F150W. We thus do not perform SED fitting to them as they have photometric data in only three bands, although we can estimate limitations for their UV luminosities.

We choose to use the double-power-law model to parametrize the star-forming history of our galaxies. We perform the SED fitting within a broad parameter space. The falling slope and the rising slope are assigned to vary in the logarithmic ranges of $-8 - 2$ and $-2 - 8$, respectively. The time scale $\tau$ (related to the SFR peak time) vary across the whole cosmic time. We assume a Calzetti law \citep{calzetti00} for dust attenuation with absolute attenuation in the V band ($A_V$) varying between 0 and 4 magnitudes. The metallicity is set within the range of 0.001 and 10 times the solar metallicity. The formed mass vary in the logarithmic range of $7-13$. We also include the Cloudy-model nebular emission with ionization parameter of $-4<{\rm log}(U)<-1$. The observed and best-fit model SEDs are shown in Figure \ref{sed}. We get proper fitting results for most of our sources except z6lae-091 and z6lae-101 (see the panels 06 and 11 in Figure \ref{sed}) which both have very red F277W$-$F444W colors. We discuss the two sources in the next section.

We obtain SFR and $M_*$ for each LAE in our sample. The SFR-$M_*$ distribution of our sample is plotted in the left panel of Figure~\ref{pps}. We also compare our results with the best-fit main sequence (MS) relations of star-forming galaxies at $z\sim3-6.5$ from \citet{rinaldi22}. Note that the evolution of the slopes is negligible in the redshift range. We can see that our LAEs lie around the MS relation and the five individual galaxies in the merging/interacting systems overall coincide with the LAEs with single components. Most of them exhibit a potentially weak-burst mode. Despite the large scatter, they populate basically in accord with the bimodal distribution indicated by the young and old LAEs \citep[e.g.,][]{2023arXiv230908515I}.

We also investigate the relation between the galaxy size and stellar mass of our sample. The galaxy size is the effective radius \Reff, which is measured from the stacked SW (F115W and F150W) images. Our results are plotted in the right panel of Figure~\ref{pps}. We compare the result of \citet{2023arXiv230706336L}. Our results overall populate below their best-fit linear function of the mass-size relation over a redshift range of $z\sim4-10$. Note that its intercept has a significant scatter of ${\sim}0.3$ dex. The sample potentially displays an overall upward trend. As our sample size limit the accurate fitting analysis, we perform linear regression fitting with a fixed slope same as that of \citet{2023arXiv230706336L}. We obtain a best-fit linear relation lower than that of \citet{2023arXiv230706336L} by ${>}0.2$ dex. The (rest-frame) UV compactness of the LAEs imply that they tend to be metal-deficient \citep[e.g.,][]{2023arXiv230706336L} and thus dust-poor, which agree with our (SED-fitting derived) dust extinction of $E(B-V)<0.1$ for our LAEs. Such ISM environment may facilitate their \lya\ escape from star-forming regions even if with bursty activities or rapid gas accretion.

\begin{figure*}[t]
\centering
\includegraphics[angle=0, width=0.85\textwidth]{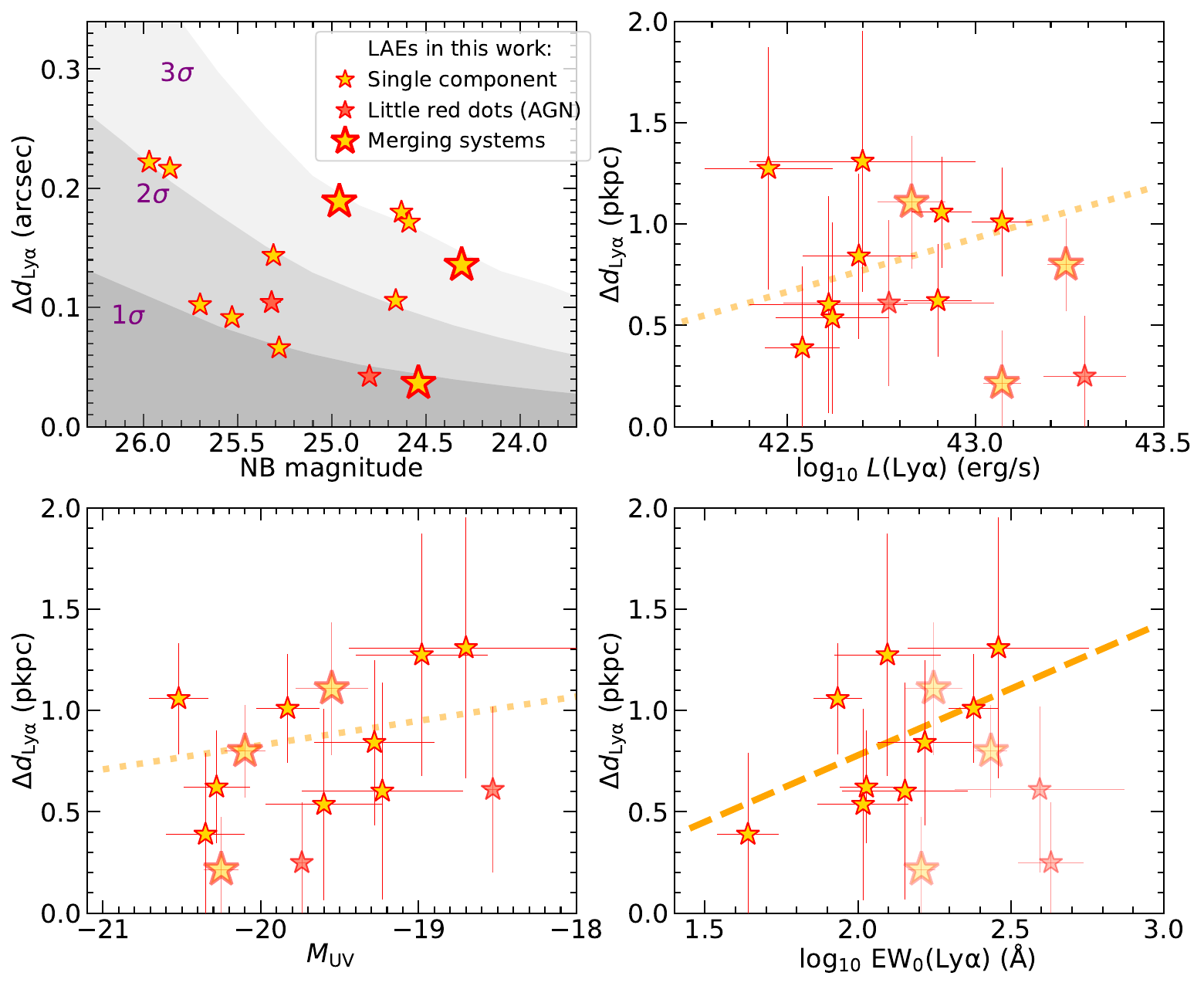}
\caption{\lya-UV spatial offset \Dd\ as a function of NB magnitude, \lya\ luminosity, rest-frame UV magnitude, and \lya\ equivalent width. The star symbols indicate the 14 LAEs $z\approx5.7$ in this work while the larger ones with thicken red edges mark the merging/interacting systems and the red-filled ones mark the two LAEs (z6lae-091 and z6lae-101) probably hosting AGNs. In the upper-left panel, the \Dd\ are plotted in unit of arcsec with the shaded regions of measurement uncertainty for the NB imaging detection. In the other panels, the \Dd\ are plotted in unit of physical kpc (pkpc) with $1\sigma$ error bars converted from the $1\sigma$ curve shown in the upper-left panel. The orange dotted and dashed lines indicate the best-fit linear functions. In the lower-right panel, we make the merger and AGN symbols shallower to highlight the LAEs used in the linear fitting.
\label{offset}}
\end{figure*}

\subsection{\lya-UV Misalignment}

We investigate the misalignment between \lya\ and UV emission for the LAEs in our sample. In this work, we compare the \lya\ centroid in the Suprime NB images and (rest-frame) UV centroid in the \jwst/NIRCam images. For the three merging/interacting systems in our sample (z6lae-072, z6lae-128, and z6lae-137), we adopt the positional center of all components as the original position. When evaluating the \lya-UV projected offsets (\Dd), we consider both accuracy and precision on the astrometry. We first address the astrometric accuracy by eliminating the overall systematic WCS offsets between the COSMOS NB image and each JWST COSMOS-Web tile. In this procedure, all detected sources in each tile are compared to their counterparts in the NB image. We thus have the median value for each two-dimensional (2d) distribution of the positional differences. The COSMOS-Web tiles have WCS systematic offsets relative to the NB image in a range of ${\sim}0\arcsec.01-0\arcsec.04$. 

For the above mentioned (2d) distributions, the dispersion degree basically reflect the astrometric precision. But it depends on the brightness and morphology of sources. In this work, we address the astrometric precision by estimating the measurement uncertainty, which mainly comes from detection measurements in the NB image. The reason is as follow.
For galaxies at $z\sim6$, a typical size of 1 kpc corresponds to ${\lesssim}0\arcsec.2$. The COSMOS NB816 image has a PSF FWHM of ${\sim}0\arcsec.5$ overwhelmingly larger than those of the NIRCam SW bands. The measurement uncertainty thus influences the positional information of source detection, especially for the faint targets. Such errors could be even comparable with their intrinsic sizes. We thus need to evaluate how the detected \lya\ centroid deviates from its original position. We implement a Monte Carlo simulation to measure the distribution of the positional offsets as a function of NB brightness. The simulation is similar to that conducted in our previous work (see Section 4.1.1 in \citealt{ning21}). 
Below we simply describe the main procedures.

We first obtain a PSF of the COSMOS NB816 image and create a mock LAE by convolving the PSF difference kernel with the stacked LAE image based on a large sample of LAEs at $z\approx5.7$ \citep{ning20}. We then simulate a large number of mock LAEs with a series of magnitudes and insert them randomly into the COSMOS NB816 image. In the meantime, we record the original positions of the mock LAEs falling on the image. Next we run \texttt{SExtractor} to detect the corresponding mock sources. For the detected ones, we compute their positional differences from their original positions. We obtain a series of 2-dimensional normal distributions of the positional offsets whose standard derivations are an increasing function as the source brightness goes fainter. The standard derivation represent the measurement error when evaluating the positional difference for the mock sources falling in the image. We finally interpolate and produce the measurement errors using the detected NB magnitude for each LAE. The results are consistent with the 2d positional difference distributions mentioned in the first paragraph of this section.

In Figure~\ref{offset}, we show the \Dd\ as a function of NB magnitudes, \lya\ luminosity, rest-frame UV magnitude, and \lya\ equivalent width. We overplot the gray shaded regions of the measurement errors from source detection in NB imaging in the upper-left panel. The results of our measured \Dd\ are listed in units of kpc in Column 4 of Table \ref{tab2}. 
For the luminous LAEs in our sample, the misalignment of \lya\ and UV emission have offsets in a range of \Dd~$\sim0-1.5$ kpc with a median value of ${\sim}0.7$ kpc. Half of our sample have large \Dd\ at a significant level of ${\gtrsim}2\sigma$. The \Dd\ are more than twice larger than the half-light radii of the LAEs on average. Previous studies also report a significant \lya-UV offset for most of \lya-emitting galaxies \citep[e.g.][]{hoag19, lemaux21, claeyssens22}. Note that the \Dd\ range of our LAE sample does not exceed a typical \lya-halo scale length of ${\sim}2$ kpc \citep{wu20}.

Our results reveals the non-negligible misalignment between the \lya\ and UV centroids for the LAEs, especially those with high \ewlya. The \lya-UV misalignment corresponds to a positional offset reaching more than 0\arcsec.2 for galaxies at $z\lesssim6$. If the redshift goes higher, for those targets of interest for JWST spectroscopic observations (while \lya\ line can fall into the JWST wavelength coverage), such angular offset becomes larger. However, for JWST/NIRSpec (multi-object) spectroscopy, the silt width (in the dispersion direction) of each MSA shutter is just 0\arcsec.2. So a significant proportion of \lya\ emission would be probably lost besides the slit losses if the shutter is aligned mainly to the UV counterparts. Such systematic offset of \lya\ emission should not be ignored when it is observed by JWST slit spectroscopy for galaxies in the epoch of reionization.

\section{Discussion}

In this section, we discuss the merger/interacting systems and probable active galactic nuclei (AGN) in our LAE sample. We also discuss the possible relation between \Dd\ and \ewlya.

\subsection{Potential Mergers in the LAE Sample}
\label{sec:pair}

In the sample, three LAEs (z6lae-072, z6lae-128, and z6lae-137) displaying galaxy pairs are potential merging/interacting systems as shown in Figure~\ref{samp}c. The number ratio (3/14) to the whole sample is similar to that of a recent study on LAEs at $z\approx3.1$ \citep[2/10;][]{2023arXiv230911559L}. Among our three double-component systems, two are at the bright end of $-42.9<$~\lgLya~$\leq-43.4$ or $-20.5<M_{\rm 1500}\leq-19.5$ making a fraction of ${\sim}30$\% while one is at the faint end of $-42.4<$~\lgLya~$\leq-42.9$ or $M_{\rm 1500}>-19.5$ giving a fraction of ${\lesssim}20$\%. The extrapolation of the two percentages is well consistent with the value of $40\%-50\%$ at a brighter range of $M_{\rm 1500}\leq-20.5$ \citep{jiang13b}. It is worth noting that two LAEs at $z\approx6.6$, Himiko and CR7 which are extremely luminous in \lya, are both triple-component systems \citep{ouchi13, sobral19b}. The result reveals a trend that the multi-component fraction goes larger as the systems are brighter in terms of \lya\ or UV which preferentially trace more overdense regions \citep[e.g.,][]{2023arXiv231104270H}.

It is interesting to see that the three LAEs are very luminous in \lya\ especially two of them are the most luminous ones. But where the \lya\ emission mainly originate from in the galaxy pairs is poorly resolved due to the relatively low resolution of NB imaging observations. It is most likely contributed by the star-forming regions because components in merger systems intuitively have higher SFRs due to triggered bursty activity. However for the three pairs, most of their components do not deviate from the star-forming MS relation as shown in the left panel of Figure~\ref{pps}. This illustrates that SFR is not apparently enhanced at the pair separation scale, which is consistent with previous results at lower redshift \citep[e.g.,][]{pearson19, daiy21}. For z6lae-137, although one of the components z6lae-137b has a weak SB mode (${\sim}0.3$ dex higher than the MS relation), the \lya\ centroid does not locate more closely to it in a close visual inspection. So the merging/interacting systems probably favor that \lya\ origin may occur in the circumgalactic medium (CGM) regions. In merger activities, strong interactions between CGM of each component can induce gas cooling \citep[e.g.,][]{sparre22, gupta23}, then produce more additional \lya\ emission.

\subsection{Little Red Dots and AGN}
\label{sec:agn}

In above SED fitting, we analyze our LAE sample assuming they are all pure galaxies. However, the best-fit model SEDs can not match the observed for two sources with the smallest galaxy sizes, z6lae-091 and z6lae-101, if we assume they are pure galaxies. We thus consider the possible existence of AGN activity which can not be ruled out for \hz\ objects \citep[e.g.,][]{juodzbalis23, 2023arXiv230906932L, 2023arXiv231012330L}. Although the typical narrow \lya\ lines in our LAE sample exclude the possibility of bright Type-1 AGNs, the broad-line component of faint AGNs may also be drowned in the noise of (ground-based) Magellan/M2FS spectra unlike much more sensitive JWST/NIRSpec. Note that they also have two highest \ewlya\ in our sample, albeit with relative large errors.

As seen in the images, z6lae-091 and z6lae-101 are both red compact sources featured as ``little red dots" \citep[e.g.,][]{2023arXiv230605448M, 2023arXiv231107483W}. Firstly, their PSF-like morphology imply the non-negligible AGN component \citep[e.g.,][]{2023arXiv230514418B}. The two LAEs are also very red in terms of the F277W$-$F444W color. They have F277W$-$F444W $\gtrsim1.6$, revealing a probability hosting AGN of at least 80\% \citep{2023arXiv230905714G}. The AGN fraction of our LAE sample is then $\frac{2}{14}\times80\%\approx0.11$, which is consistent with that faint AGNs occupy ${\sim}10\%$ of galaxy population at $z\sim6$ \citep[e.g.,][]{harikane23, 2023arXiv230801230M}.

Although a strong \ha\ line may also boost the F444W band for $z\sim6$ galaxies, it is worthy noting that a bluer F277W$-$F444W color of ${<}1.5$ is shown by a LBG \citep[from][]{ning22} with an extremely high ionizing production efficiency of log$_{10}$ \ksio\ (Hz erg$^{-1}$) $\approx26.5$ which is a rare case in \hz\ galaxies \citep[e.g.,][]{flares13, 2023arXiv230915671R}. In addition, the ALMA ALPINE survey has covered z6lae-101 in the band~7 \citep{bethermin20}. The non-detections in both \cii\ 158$\mu$m line and far-IR continuum ($3\sigma$ upper limits of 0.190~Jy~km/s and 0.194~mJy, respectively) also disfavor that the dusty star formation results in such red optical continuum. So pure galaxies (at $z\sim6$) should hardly produce the very red F277W$-$F444W color. The AGN presence in the LAE population waits to be revealed with rest-frame optical spectra or images in other bands (such as MIRI) based on a larger sample.

\subsection{\Dd-\ewlya\ Relation}
\label{sec:dewlya}

\lya\ emission provides useful information for probing reionization process and physics of galaxies in this epoch. However, \lya\ photons propagate with resonant scattering through complicated routines and suffer the absorption in the interstellar medium (ISM), CGM and the surrounding IGM (e.g. \citealt{hayes11, dijkstra14, cai17a, cai19, zhangh24}). It is thus not surprising that \lya\ emission exhibits at different positions from the star-forming regions of galaxies. In the meanwhile, many studies utilize such \lya-UV (projected) offsets in turn to explore \lya\ escaping process. For example, \citet{shibuya14a} show that some LAEs (at $z\sim2.2$) have a spatial offset reaching ${\sim}2.5-4$ kpc between the \lya\ and UV-continuum emission.

In Figure~\ref{offset}, the larger star symbols mark the merging/interacting systems while the red-filled ones indicate the two LAEs, z6lae-091 and z6lae-101, probably hosting AGNs. For the merging/interacting systems, the \lya\ centroid is supposed to be modified by all components, which is a complex process. For the (Type-2) AGNs, \lya\ emission is also radiated from the narrow-line region \citep{agn2} besides the star-forming region of galaxies. If we ignore the LAEs hosting AGN or the components in the lower-right panel, we can see that there is a likely positive correlation between \Dd\ and \ewlya\ with a fitting linear slope of $0.66\pm0.37$. The correlations become weaker in the \Dd-\muv\ and \Dd-\Llya\ parameter spaces. The Spearman correlation coefficients are $0.34\pm0.24$, $0.18\pm0.26$, and $0.12\pm0.32$ for the \Dd-\ewlya, \Dd-\muv, and \Dd-\Llya\ relations, respectively, by perturbing the measured quantities with their uncertainties \citep[e.g.,][]{curran14}. Note that the original coefficient 0.57 for \Dd-\ewlya\ is largely reduced after perturbing the dataset due to the small sample size and large uncertainty.

The \Dd-\muv\ trend does not agree with \citet{lemaux21}. They found UV-brighter galaxies exhibit offsets ${\sim}3$ times larger than their UV-fainter counterparts. The reason for such inconsistency is unclear, but we notice that they measure the \lya\ location based on the two-dimensional spectra of \lya\ lines through the slits. One of the possibilities might be that our LAEs are stronger in terms of \ewlya\ while most of their sample have \ewlya~$<100$ \AA. If we extrapolate our \Dd-\muv\ trend into a range of \muv~$\lesssim-22$ mag, the \Dd\ of LAEs is supposed to be smaller, which is also consistent with \citet{jiang13b}. They find that the observed \lya\ location does not deviate from its original position in compact galaxies based on a UV-brighter sample.

Here we invoke a simple physical scenario to explain the marginal \Dd-\ewlya\ relation. A basic picture is wrapped up inside the diffusion process of \lya\ transfer in ISM (and/or CGM) \citep{dijkstra14, gronke16}. \hi\ density or its covering fraction plays an important role to modify both \Dd\ and \ewlya. If the \hi\ gas is denser (or with higher covering fraction), \lya\ emission will escape more hardly from the original region (smaller \Dd) while the metallicity (and thus dust content; e.g., \citealt{popping17}) is correspondingly higher \citep{santini14, tgs20} inducing lower \ewlya\ (also lower \lya\ escape fraction \fescLya; e.g., \citealt{roy23}). Conversely, higher \ewlya\ emission can wander off (anisotropically) more easily with larger \Dd\ \citep{df18} in the less chemically enriched ISM \citep[e.g.,][]{maseda23, 2023arXiv230604536S}. This scenario can also explain the previously found anti-correlation between \lya\ velocity offset \Dv\ and \ewlya\ \citep{tangm21a, lyon23}. As \lya\ photons are scattered by \hi, the offset in real (\Dd) and frequency (\Dv) space couple with each other. The ionized bubble size can thus be estimated by large \Dd\ misalignment which is associated with low \Dv\ \citep[e.g.,][]{mason20} for the galaxies in the epoch of reionization. The above scenario can also be tested with additional \fescLya\ results based on more spectroscopic information including Balmer lines such as \ha.

\section{Summary}
In this work, we study a sample of LAEs at $z\approx5.7$ based on JWST/NIRCam imaging data. The sample consists of 14 LAEs which have been spectroscopically confirmed with strong \lya\ lines previously by our Magellan/M2FS survey. They spread a \lya\ luminosity range of \Llya~$\sim10^{42.4-43.4}$~erg~s$^{-1}$. The JWST/NIRCam imaging dataset comes from the COSMOS-Web program which is still ongoing currently but has covered around half of total designed survey area in the four NIRCam bands (F115W, F150W, F277W, and F444W). Our LAEs constitute a sample of luminous galaxies in terms of \lya\ luminosity over a large survey area in this redshift slice.

We utilize the highly resolved, deep near- and mid-IR images to explore a variety of galaxy properties including the rest-frame UV properties (luminosity, continuum slope, and size), stellar mass ($M_*$), star-formation rates (SFRs). We also investigate the misalignment of \lya\ and UV emission for the LAEs at $z\approx5.7$. We summarize our conclusions as follows:

\begin{itemize}
 \item{The LAE sample are UV-faint between \muv~$\sim{-}20.5$ and ${-}18.5$~mag. The UV-continuum slopes have a median value of \buv~$\approx-2.35$ in a range from ${-}3.5$ to ${-}1.0$. These luminous LAEs in the sample are overall bluer. A fraction (6/15) of them (including the components) exhibit very blue UV slopes of \buv~$<-3$ with a mean $-3.46\pm0.43$.
 }
 \item{With the galaxy SED modeling, we derive stellar mass and star-formation rates for the whole sample. Most of LAEs or their components lie on the main-sequence relation while a small fraction of them show a weak-burst mode.
 }
 \item{Our LAEs are overall compact in morphology. Besides the two point-like sources, the galaxy size \Reff\ are between 0.1 kpc and 0.7 kpc with a median \Reff\ of ${\approx}0.2$ kpc. The LAEs are found substantially (${>}0.2$ dex) below the overall mass-size relation of high-redshift galaxies.
 }
 \item{Three of the 14 LAEs are distinguished as potential merging/interacting systems. Estimating the merger fraction at the bright and faint bins indicates that luminous LAEs (in UV or \lya), which preferentially reside in more overdense regions, tend to be interacting systems. The gas cooling induced by the interaction may enhance \lya\ luminoisity.
 }
 \item{Two LAEs in our sample probably own non-negligible AGN components. Both of them are featured as ``little red dots" with the PSF-like morphology and very red color in F277W$-$F444W. The probability of our sample containing AGN is roughly $10\%$, which is consistent with the AGN fractions obtained by previous works.
 }
 \item{For our luminous sample of LAEs at redshift 5.7, the misalignment between \lya\ and UV emission mostly exist with positional offsets \Dd\ in a range of ${\sim}0-1.5$~kpc (a median value of ${\sim}0.7$~kpc). A positive correlation may exist between \Dd\ and \ewlya, which could be established by varying \hi\ density (and/or covering fraction) of ISM and CGM. Such \lya-UV offsets should be considered when conducting JWST slit spectroscopic observations if covering \lya.
 }
\end{itemize}

Our results reveal a diversity of the sources associated with LAEs, including AGNs, merging/interacting systems, and single star-forming galaxies. This work also highlights the powerful performance of JWST to study luminous LAEs at redshift $z\gtrsim6$ over a relatively large sky area (${\sim}0.28$ deg$^2$). We will extend our research by enlarging the sample and including the $z\approx6.6$ LAEs confirmed by our spectroscopic survey with the entire dataset from COSMOS-Web survey. We also look forward to unravel the nature of LAEs in the epoch of reionization with more JWST spectroscopic followup and imaging observations in the other bands.

\acknowledgments
{We thank the COSMOS-Web group lead by J.~S.~Kartaltepe and C.~M.~Casey for contributing the great treasury JWST imaging program.
We thank L.~Jiang and J.~Li for the instructive discussions. We thank Z.~Sun for maintaining the high-performance computing platform of the high-redshift research group in Department of Astronomy, Tsinghua University. We also thank the anonymous referee for the constructive comments and suggestions that improved this paper.

We acknowledge support from the National Key R\&D Program of China (grant no.~2018YFA0404503), the National Science Foundation of China (grant no.~12073014), the science research grants from the China Manned Space Project with No.~CMS-CSST-2021-A05, and Tsinghua University Initiative Scientific Research Program (No.~20223080023). Q. Li acknowledges support from the ERC Advanced Investigator Grant EPOCHS (788113).

This work is based on the observations made with NASA/ESA/CSA James Webb Space Telescope. \jwst\ data are obtained from the Mikulski Archive for Space Telescopes (MAST) at the Space Telescope Science Institute, which is operated by the Association of Universities for Research in Astronomy, Inc., under NASA contract NAS 5-03127 for \jwst. The \jwst\ observations are associated with the program GO-1727.}

\facilities{\jwst\ (NIRCam)}
\bibliography{ms.bbl}
\end{document}